\documentclass[english,aps,prc,0superscriptaddress,longbibliography,notitlepage]{revtex4-1}
\usepackage{amsmath,amssymb,multirow,bm}
\usepackage{soul}

\usepackage{epsfig}
\usepackage{graphicx}
\usepackage{amsmath}
\usepackage{lipsum}
\usepackage{color}
\makeatother

\usepackage{babel}
\usepackage[colorlinks,linkcolor=blue,anchorcolor=blue,citecolor=blue,urlcolor=blue,filecolor=black]{hyperref}

\newcommand{\beq}{\begin{equation}}
\newcommand{\eeq}{\end{equation}}
\newcommand{\bea}{\begin{eqnarray}}
\newcommand{\eea}{\end{eqnarray}}
\usepackage{graphicx}
\preprint{}

\begin{document}


\title{Nuclear Fragmentation at the Future Electron-Ion Collider}
\author{C. A. Bertulani}
\email{carlos.bertulani@tamuc.edu}
\affiliation{Department of Physics and Astronomy, East Texas A\&M University, 
Commerce, TX 75429, USA }
\affiliation{Institut f\"ur Kernphysik,  Technische Universit\"at Darmstadt, 64289 
Darmstadt, Germany}
\affiliation{Helmholtz Research Academy Hesse for FAIR, D–64289 Darmstadt, Germany}

\author{Y. Kucuk}
\email{ykucuk@akdeniz.edu.tr}
\affiliation{Turkish Accelerator and Radiation Laboratory (TARLA), 06830, Ankara,
Turkey}
\affiliation{Akdeniz University, Department of Physics, 07058, Antalya, Turkey}

\author{F. S.  Navarra}
\email{navarra@if.usp.br}
\affiliation{Instituto de F\'\i sica, Universidade de S\~ao Paulo, 
Rua do Mat\~ao 1371, CEP 05508-090, Cidade Universit\'aria, S\~ao Paulo,Brazil}

\date{\today}

\begin{abstract}
We explore the potential of conducting low-energy nuclear physics studies, including nuclear structure and decay, at the future Electron-Ion Collider (EIC) at Brookhaven. 
By comparing the standard theory of electron-nucleus scattering with the equivalent
photon method applied to Ultraperipheral Collisions (UPC) at the Large Hadron 
Collider (LHC) at CERN. In the 
limit of extremely high beam energies and small energy transfers, very transparent  
equations emerge. We apply these equations to analyze nuclear fragmentation in 
UPCs at  the LHC and $eA$ scattering at the EIC, demonstrating that the  EIC could facilitate 
unique photonuclear physics studies. However, we have also shown that the fragmentation cross-sections at the EIC are about 1,000 times smaller than those at the LHC.  At the LHC, the fragmentation of uranium nuclei displays characteristic double-hump mass distributions from fission events, while at the EIC, fragmentation is dominated by neutron emission and fewer few fission products, about 10,000 smaller number of events.

\end{abstract}
 

\maketitle

{\it Introduction}.
The upcoming Electron-Ion Collider (EIC) at Brookhaven National Laboratory (BNL) 
will address fundamental questions in Quantum Chromodynamics (QCD), such as the 
role of sea quarks and gluons, their spin contributions, and their spatial and 
momentum distributions within the nucleon and the nucleus. The nuclear environment is expected 
to influence these distributions and gluon interactions within nuclei    \cite{accardi2014}. Additionally, there is significant anticipation that  
low-energy nuclear physics experiments can be conducted at the EIC, where     
the fragmentation of nuclei in electron-ion collisions could produce numerous 
isotopes, including exotic and potentially undiscovered rare isotopes \cite{magdy2024,Schmookler2023}. Low-energy nuclear spectroscopic studies may  
also be feasible by detecting photons in the far-forward detection area, where 
Doppler-shifted photons can reach high energies. The resulting time dilation 
effect enables the detection of lifetimes as short as a few nanoseconds, because 
high-energy photons are easier to detect.

Fragments produced in  Coulomb excitation of relativistic nuclei in ultra-peripheral collisions (UPC) have provided relevant information on fission dynamics, the discovery of new isotopes and isomeric states, and more than 1000 nuclear fission residues \cite{SCHMIDT2002157,JURADO2003186,JURADO200514}. Therefore, relativistic Coulomb excitation provides an extremely useful tool for studying the pathway to fission, studies of neutron emission and their relation to the astrophysical rapid capture (r-)process and to many other nuclear decay studies \cite{PEREZLOUREIRO2011552}. The fragmentation process proceeds via the excitation of giant resonances and in particular the nuclear dipole response is probed reflecting the isospin imbalance in the nucleus. This can be used to  constrain the slope of the symmetry energy $L$ of the nuclear matter equation of state (EoS) \cite{ROCAMAZA201896,TamiPRL.107.062502}. As an alternative to UPCs in relativistic heavy ion 
collisions it is worthwhile investigating 
how the future EIC can be used to study the dynamics of 
excitation and decay and if new nuclear isotopes can be produced that are not accessible in heavy ion collisions.  This is the main motivation for this work focusing on the specific   
issue of nuclear fragmentation due to the electromagnetic (EM) interaction of 
electrons with nuclei at low energies ($\hbar \omega < 50$ MeV).

The most significant region for nuclear fragmentation induced by real photons is 
within the giant dipole resonance (GDR) region, with excitation energies around 
10-20 MeV for heavy nuclei, and slightly higher for lighter nuclei.           
Electron-nuclear interactions have a large cross-section for exciting GDRs or 
high-energy nuclear states. These GDRs typically decay by emitting neutrons,   
protons, light elements, and, in the case of actinides, fission fragments. The     
neutrons emitted in this process have low energies in the ion frame, significantly 
lower than the energy scales of neutrons produced in central relativistic heavy-ion
collisions. As a result, these emitted neutrons can serve as indicators of strong  
electromagnetic fields. In the laboratory frame, these neutrons will have an energy 
comparable to the ion energy per nucleon.

In this study, we demonstrate that neutron emission and fission, primarily driven 
by EM dissociation of nuclei through GDR decay, are highly significant and may be  
observed abundantly at the future EIC. We also compare the absolute yield of  
fragments at the EIC with those in UPCs of heavy 
ions at the Large Hadron Collider (LHC) at CERN \cite{BERTULANI1988299}. Our     
results highlight the role of different multipolarities and the correspondence 
between virtual photons at the EIC and UPCs and pinpoint the key features that 
distinguish low-energy photonuclear physics in these two processes. Throughout 
this work, we use natural units with $\hbar = c = 1$.

\medskip
{\it Electron scattering at ultra-high energies.} In electron-nucleus scattering, 
the momentum transfer $Q^2$ in terms of the electron energy $E$ and scattering 
angle $\theta$ is given by
\beq
Q^2 = -(k-k')^2 \simeq 2EE'(1-\cos \theta) \simeq 4E^2\sin^2(\theta/2),  \label{Q2}
\eeq
where the energy and momentum  transfer are $\omega = E - E'$ and 
${\bf q} = {\bf k} - {\bf k}'$, respectively, so that           
$Q^2 = q^2+\omega^2$. It is easy to show that, for a given energy transfer 
$\omega$, the minimum momentum transfer is $Q_{\text{min}} = m_e\omega/E$, 
with $m_e$ being the electron mass. 

In 1924, Fermi demonstrated that a high energy electric charge passing near a atom at a 
distance $b$ (impact parameter) can be analyzed by Fourier  
transforming the field it generates and relating it to a flux of equivalent real 
photons with energy $\omega$ \cite{Fermi1924,Fermi1925}. This concept was later 
extended by Weizs\"acker and Williams (WW) \cite{Weisz34,PhysRev.45.729} and the following expression for the number of equivalent photons emerges, integrated 
over collisions at all impact parameters:
\beq
n(\omega) = \frac{2Z^2e^2}{\pi \omega} \left[ xK_0(x) K_1(x) - x^2 \left(K_1^2(x) 
- K_0^2(x)\right)\right] 
= \frac{2Z^2e^2}{\pi \omega}\ln(x), \label{eparhic}
\eeq
where $Z$ is the particle charge, $\omega$ is the photon energy, and    
$x = \gamma / \omega b_{min}$, with $\gamma = (1-v^2)^{-1/2}$ being the    
Lorentz contraction factor for a projectile with velocity $v\sim c$, and      
$b_{min}$ being the minimum impact parameter. This expression matches Fermi's 
result for non-relativistic energies by just using $\gamma=1$.  
The rightmost formula is particularly accurate for low-energy photonuclear 
studies at the LHC.

Nordheim, Nordheim, Oppenheimer, and Serber \cite{PhysRev.51.1037} have shown 
that when applying this expression to electron scattering, the smallest impact 
parameter $b_{\text{min}}$ should be taken as $1/k_{\text{max}}$, where 
$k_{\text{max}}$ is the maximum effective momentum transfer. While this momentum 
transfer can be as large as the electron energy, a more appropriate value for 
$k_{\text{max}}$ is $1/R$, where $R$ is the nuclear radius. Various studies, some 
using the extended WW formula and others based on the quantum electrodynamics (QED) 
treatment of electron-nucleus scattering, have yielded slightly different 
expressions for the number of equivalent photons  \cite{PhysRev.105.1598,ISABELLE1963209,BUDNEV1975181}.
Here, we will derive accurate expressions for the equivalent photon numbers directly 
from the conventional electron scattering formulas. Additionally, we will show that 
certain terms, which are crucial for electron-nucleus scattering at lower electron 
energies, become negligible when the electron energy significantly exceeds the 
energy-momentum transfer to the nucleus.

In the Plane Wave Born Approximation (PWBA), the electron-nucleus cross section is 
given by \cite{ForestWalecka1966,DonnellyWalecka1975,eisenberg1988excitation}
\beq
\frac{d\sigma}{d\Omega d\omega} = 4\pi \sigma_M f_{\text{rec}} \sum_\lambda             
\left[ {Q^4\over q^4} \left| F_\lambda^L(q) \right|^2 + \left( \frac{Q^2}{2q^2} 
+ \tan^2 \frac{\theta}{2} \right) \left| F_\lambda^T (q)\right|^2 + 
\left| F_\lambda^M(q) \right|^2 \right] g(\omega)\, , \label{rcc}
\eeq
where $\lambda \geq 1$ denotes the multipolarity, $F_\lambda^L$, $F_\lambda^T$, and 
$F_\lambda^M$ are the Coulomb longitudinal, electric transverse, and magnetic form 
factors, respectively. $g(\omega)$ is the density of states in the continuum as we will apply this equation to
the excitation of giant resonances. The Mott cross section is 
\begin{equation}
\sigma_M=\alpha^2{\cos^2(\theta/2)\over 4E^2\sin^4(\theta/2)}=4\alpha^2
\left(1-{Q^2\over 4E^2}\right){E^2\over Q^4},
\end{equation} 
where we used $Q^2=4E^2\sin^2(\theta/2)$,  $\alpha=e^2$ is the fine-structure 
constant, and $e$ is the elementary charge. The nuclear recoil correction is
$
f_{rec}= \left( 1 +2E\sin^2(\theta/2)/M\right)^{-1}
$
where $M$ is the nuclear mass. For the energies at the future EIC, 
$f_{rec} \simeq 1$ because $E \gg M$, although for large angle scattering it may become significant.

The Coulomb longitudinal form factors for the nuclear transition from initial to 
final states are given by
\beq
eF_\lambda^L(q,\omega) = \frac{\hat{J_f}}{\hat{J_i}} \int_0^\infty \delta 
\rho_\lambda (r,\omega) j_\lambda (qr) r^2 \, dr, \label{efq1}
\eeq
where $\hat{J} = \sqrt{2J+1}$, with $J_i$ and $J_f$ being the initial and final     
angular momenta of the nucleus, respectively. Here, $j_\lambda(x)$ is the spherical 
Bessel function, and the transition density is
$
\delta \rho_\lambda = \left< \Psi_f \left| \hat{\rho}(\mathbf{r}) Y_\lambda \right| 
\Psi_i \right>,
$
with $\hat{\rho}$ being the electric charge operator and $Y_\lambda$ the spherical   
harmonics. Similar expressions for the electric transverse and magnetic form factors 
involving transition currents can be found in the literature \cite{ForestWalecka1966,DonnellyWalecka1975,eisenberg1988excitation}. 
The photonuclear cross section encodes nuclear structure information, including  
details about nuclear wavefunctions. We will use known experimental photonuclear 
scattering data on giant resonances to make predictions, thereby avoiding explicit use of nuclear wavefunctions. At energies higher than 
the giant resonances, sub-nucleon degrees of freedom become relevant, and       
first-order perturbation theory, as implied by Eq. \eqref{rcc}, may not suffice 
for modeling the reaction accurately.

When one applies Eq. \eqref{rcc} to study electron scattering, including angular   
distributions, a few more corrections are in order. The Coulomb attraction between 
the electron and the nucleus causes the electron to accelerate as it approaches the 
nucleus, leading to a focusing effect that concentrates the electron wavefunction 
more onto the nucleus. In their interaction with the nucleus, electrons ``see" a 
higher momentum transfer. The PWBA form factors should be modified so 
that $ Q $ is shifted to an effective momentum transfer \cite{ForestWalecka1966,DonnellyWalecka1975,eisenberg1988excitation}:
$
Q_{eff} = Q \left( 1 + {3Ze^2}/{2 E R} \right),
$
where $ R \simeq 1.12 A^{1/3} $ fm is the nuclear radius. This focusing effect 
leads to an increase in the scattering cross section and a smearing that fills in 
the minima that would be observed in a PWBA calculation. For the energies available 
at the future EIC, $E \gg Z/R$, and this correction is again insignificant. 

Collective excitations in nuclei are dominated by the excitation of giant electric 
dipole (E1) and quadrupole (E2) resonances, which decay via the emission of 
neutrons, protons, and $\alpha$-particles, or by nuclear fission. For these    
processes, the small argument approximation for the spherical Bessel function in 
Eq. \eqref{efq1} yields \cite{ForestWalecka1966,DonnellyWalecka1975,eisenberg1988excitation}, 
\begin{equation}
eF_\lambda^L(q^2) = - \sqrt{\frac{\lambda}{\lambda + 1}} \frac{q}{\omega}    
eF_\lambda^T(q^2)
\simeq - \frac{q^\lambda}{(2\lambda + 1)!!} \left[ 1 - \frac{q^2 R^2_{tr}}{2 
(2\lambda + 3)} + \cdots \right] \sqrt{B(E\lambda, \omega)},
\label{long}
\end{equation}
where $ R^2_{tr} $ is the ``transition radius'' associated with the electric       
transition density $ \delta \rho_\lambda(r) $ for the multipolarity $ \lambda $. 
The transition density is surface-dominated, implying $ R_{tr} \sim R $. In 
Eq. \eqref{long},
\begin{equation}
B(E\lambda, \omega) = \left[ \frac{\hat{J}_f}{\hat{J}_i} \int_0^\infty \delta 
\rho_\lambda(r, \omega) r^{\lambda + 2} \, dr \right]^2, \label{bel}
\end{equation}
is the reduced transition probability of the nucleus for the electric 
multipolarity $ \lambda $ and excitation energy $\omega$.

Using $d\Omega =\pi dQ^2/E^2$ in Eq. \eqref{rcc}, the total electron-nucleus 
differential cross section in PWBA can be expressed as
\begin{equation}
\frac{d\sigma}{d\omega dQ^2} = 16\pi^2\alpha {1-{Q^2/ 4E^2}\over Q^4}\sum_\lambda 
\left\{ {Q^4\over q^4} +{\lambda +1 \over \lambda}{\omega^2\over q^2} 
\left[ {Q^2\over 2q^2} +{Q^2/4E^2 \over {1-Q^2/4E^2}}\right] \right\} 
{q^{2\lambda} \over [(2\lambda +1)!!]^2}B(E\lambda, \omega) g(\omega). \label{pwba2}
\end{equation}

Low energy excitation $\omega \lesssim 50$ MeV at the EIC will  lead to very   
forward scattering, such that we can use $Q\ll E$, and to leading order in $Q$ 
the above expression reduces to
\begin{equation}
\frac{d\sigma}{d\omega dQ^2} = 8\pi^2\alpha{\omega^2 \over q^4Q^2}            
\sum_\lambda{\lambda +1 \over \lambda}{q^{2\lambda} \over [(2\lambda +1)!!]^2}
B(E\lambda, \omega) g(\omega). \label{pwba3}
\end{equation}
The cross section  induced by a real photon with energy $\omega$ and multipolarity 
$\lambda$ is known to be related to $B(E\lambda, \omega)$ as \cite{BlattWeisskopf}:
\begin{equation}
\sigma_\gamma^{(\lambda)} (\omega) = (2\pi)^3 {(\lambda + 1)\over 
\lambda [(2\lambda + 1)!!]^2} \omega^{2\lambda - 1} B(E\lambda,\omega) g(\omega) ,
\end{equation}
where the total photonuclear cross section is a sum over all multipolarities, $\sigma_\gamma (\omega) = 
\sum_\lambda \sigma_\gamma^{(\lambda)} (\omega)$. 

We can cast Eq. \eqref{pwba3} in the form 
\begin{equation}
\frac{d\sigma}{d\omega dQ^2} = \sum_\lambda \frac{dN_\lambda(\omega)}{d\omega dQ^2} 
\sigma_\gamma^{(\lambda)} (\omega), \label{epn}
\end{equation}
where the equivalent photon spectrum for a given momentum and energy transfer and 
for a given multipolarity $\lambda$ is
\begin{equation}
\frac{dN_\lambda}{d\omega dQ^2} = \frac{\alpha}{ \pi} {1 \over \omega Q^2}
\left({q\over \omega}\right)^{2(\lambda-2)}. \label{dnl}
\end{equation}
This is our main expression, derived in an extremely straightforward fashion 
 from well known electron scattering theory. It shows that at 
the EIC the electron-nucleus scattering cross section is directly proportional to the   
corresponding photonuclear cross section. This is useful as many photonuclear cross 
sections are known from experiments performed with real photons.  The dependence on 
$Q$ is very simple and the cross sections reach their maximum at the minimum     
momentum transfer $Q_{mim}=m_e\omega/E$ which is extremely small at the EIC even   
for $\omega \sim 50$ MeV. A dependence on the multipolarity also remains and it can
be useful to disentangle nuclear structure features in future experiments at the EIC 
via the analysis of the angular dependence of the nuclear excitation cross sections. 
This is specially true for  small momentum transfers, 
$q^2/\omega^2 \sim 1$ because $q^2 = Q^2+\omega^2$.

Since $dQ^2 = 2QdQ$, the integration of the virtual photon numbers  from $Q_{min}$  
to a maximum value $Q_{max}$, yields 
\begin{equation}
\frac{dN_\lambda}{d\omega} = \frac{2\alpha}{\pi} \frac{1}{\omega} \ln\left( 
\frac{EQ_{max}}{m\omega} \right), \label{dnL}
\end{equation}
which is independent of $\lambda$, so that 
\begin{equation}
\frac{d\sigma}{d\omega} = \frac{dN_\lambda(\omega)}{d\omega} \sigma_\gamma (\omega). 
\label{ep2n}
\end{equation}

The substitutions $E/m_e \rightarrow \gamma$ and $Q_{max} \rightarrow 1/R$ result  
in the same Fermi's equivalent photon equation \eqref{eparhic}. Except that the 
nuclear radius  $R$ now takes the role of the minimum impact parameter $b_{min}$ in heavy 
ion collisions. For a typical nuclear radius of $R=5$ fm, the maximum momentum 
transfer is $Q_{max} \sim 40$ MeV.  This value is right at the high energy tail 
of the giant resonances. Therefore, taking $Q_{max} =1/R$ or $Q_{max} =        
\omega_{max}$, where $\omega_{max}$ is the maximum excitation energy, does not 
make much difference as they enter in the argument of a logarithmic function. 
For excitation energies larger than $\omega_{max}$, the photonuclear cross section 
for heavy nuclei is only a few millibarns, whereas in the giant resonance region 
it can reach values close to 1 barn. $Q_{max} = \omega_{max}$ is also a good    
choice if one wants to study  particle production, such as $\Lambda$ or $J/\Psi$ 
particles. In this case, one can set $Q_{max} = M_\Lambda, \ M_{J/\Psi}$, or 
another appropriate mass for the problem under scrutiny.

The Lorentz factor entering Eq. \eqref{eparhic} needs to be calculated  in the 
frame of reference of one of the ions and  is related to the laboratory factor 
$\gamma_{lab}$ for the ions by $\gamma = 2 \gamma_{lab}^2 - 1$.  At the LHC  one 
finds $\gamma = 1.73 \times 10^7$. At the EIC, the effective electron-nucleon      
center-of-mass energy is $E = 2 \sqrt{E_{e} E_{ion}}$, where $E_e$ is the electron 
energy and $E_{ion}$ is the energy per nucleon of the ion. This yields a        
center-of-mass energy range of 45-85 GeV for electrons with energy 5-18 GeV and  
ions with energy 100 GeV/nucleon. If we use $E=80$ GeV, we find                
$\gamma =1.6 \times 10^5$ at the EIC. Therefore, the logarithmic factor in the 
equivalent photon expressions are not very different for either laboratory. But 
there is a notable difference in the magnitude of the corresponding equivalent 
photon numbers because of the factor $Z^2$ missing in the expression for the EIC. 
The cross sections for the excitation of giant resonances are therefore at least 
a factor $10^3$ smaller at the EIC compared to the LHC. 

\medskip
{\it EM Fragmentation at the LHC and at the EIC}.
To calculate the excitation of the isovector giant dipole resonance (IVGDR) and the 
isoscalar and isovector giant quadrupole resonances (ISGQR and IVGQR)  \cite{BERTULANI1988299,aumann1995ZP,BERTULANI1999139,PhysRevLett.124.132301} we 
assume a Lorentzian shape for each resonance:
\begin{equation}
\sigma_\gamma^{GR}(\omega) = \sigma_0 \frac{\omega^2 \Gamma^2}{(\omega^2 - 
\omega_{0}^2)^2 + \omega^2 \Gamma^2}.
\end{equation}
For the IVGDR, we use the energy centroid given by $\omega_{0} = 31.2A^{-1/3} + 
20.6A^{-1/6}$ MeV and its strength $\sigma_0$ determined by the                
Thomas-Reiche-Kuhn (TRK) sum rule, which is a nearly model-independent result for   
the nuclear response by an electric dipole operator \cite{eisenberg1988excitation}. 
The energy centroids of the ISGQR and IVGQR states are taken as                    
$\omega_0= 62A^{-1/3}$ MeV and $130A^{-1/3}$ MeV, respectively. All resonances 
are assumed to exhaust their associated operator sum rules, as found, e.g., in 
Ref. \cite{Harakeh:02}.
 \begin{figure}[h]
\begin{center}
\includegraphics[scale=0.39]{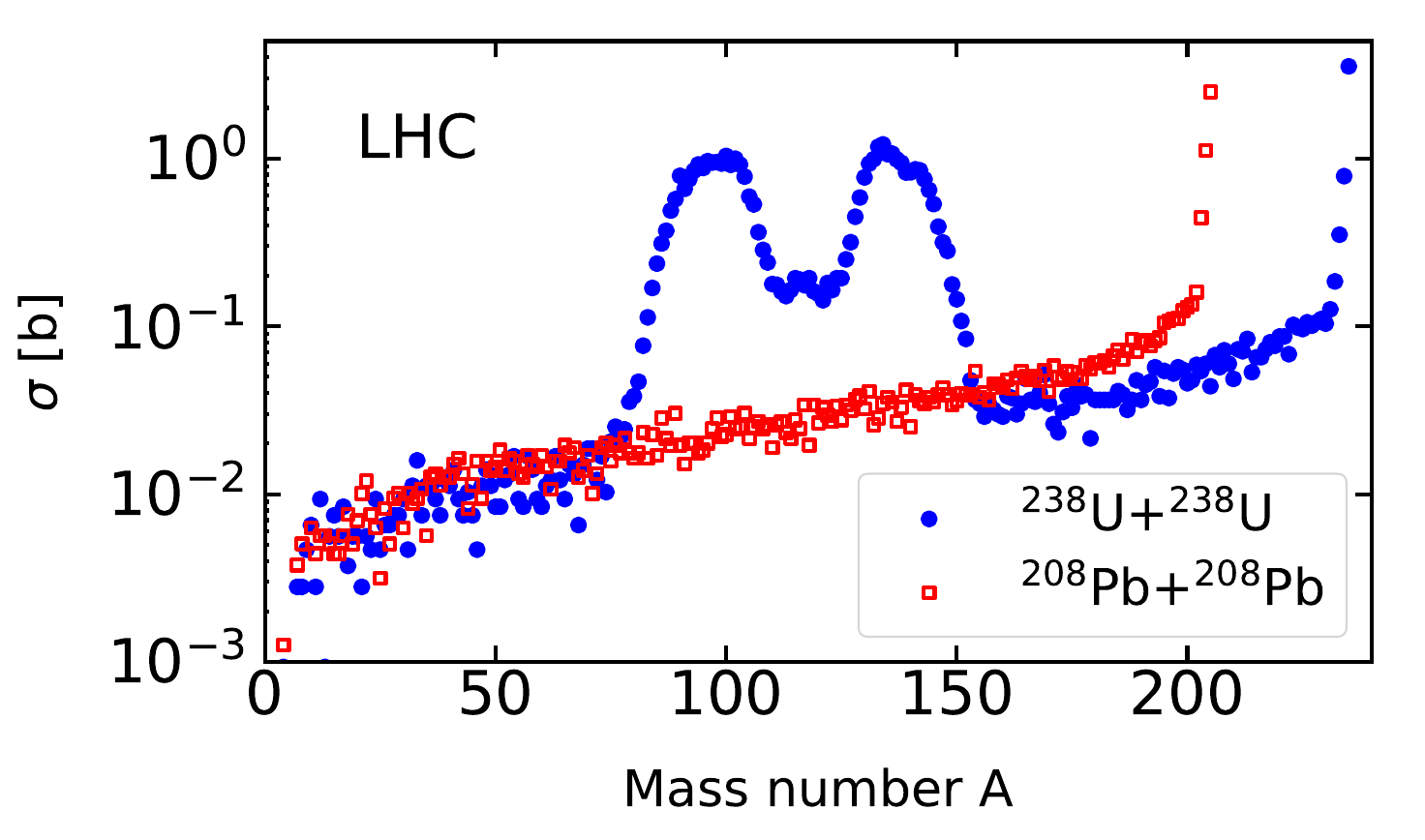}
\caption{Cross sections for the production of fragments as a function of their  
masses in ultraperipheral collisions of  $^{208}$Pb (open red squares) and 
$^{238}$U (closed blue circles) nuclei at the LHC.  \label{LHC1}}
\end{center}
\end{figure} 

We consider an 18 GeV electron beam colliding with a 110 GeV/nucleon    
${}^{238}\text{U}$ or ${}^{208}\text{Pb}$ beam at the EIC corresponding to $E_{lab} =89$ GeV, and           
${}^{208}\text{Pb} + {}^{208}\text{Pb}$ collision at the LHC with energy $\sqrt{s_{NN}} = 5.5$ TeV/nucleon, corresponding to a laboratory beam energy of $E_{lab} = 2.76$ TeV/nucleon and to a Lorentz factor $\gamma_{lab} = 2941$. 
We will also 
consider a hypothetical collision of ${}^{238}\text{U} + {}^{238}\text{U}$   
at the LHC as we want to assess the fragments arising from fission channels in 
both accelerators. 
The de-excitation of the giant resonances leads to nucleon emission, light-charged  
particles, photon emission, intermediate-mass fragments (IMFs), as well as fission 
products. The production of an isotope $x = (Z, A)$ is obtained from
\begin{equation}
\sigma_x(\omega) = b_x(\omega) \sum_{GR} \sigma_\gamma^{GR}(\omega),
\end{equation}
where the sum is carried over all giant resonances (GRs) and $b_x(\omega) = 
{\Gamma_x(\omega^*)}/{\Gamma_{tot}(\omega^*)}$ is the branching ratio or       
probability of fragment $x$ emission when the nucleus is excited to an energy 
$\omega$, where $\Gamma_x$ is the partial decay width for emitting particle 
$x$ and $\omega^*$ is the excitation energy of the compound nucleus. For the 
excitation of a giant resonance, the pre-equilibrium emissions are small, and the 
excitation energy $\omega^*$ is approximately equal to the photon energy $\omega$.
To calculate $b_x(\omega)$, we adopt the Ewing-Weisskopf model and use the ABLA07 
code \cite{aleks2009abla07}. We include the separation energies and emission 
barriers for charged particles using the 2017 atomic mass evaluation \cite{ame2016a} and the Bass potential \cite{bass} for the calculation of     
transmission probabilities. Additionally, fission yields are obtained within    
the dynamical model of Refs. \cite{JURADO2003186,JURADO200514}. ABLA07 has been  
extensively utilized in numerous publications and shown to work very well to    
describe  isotopic distributions of fission fragments measured in spallation 
and fragmentation reactions with relativistic nuclei.

In Fig. \ref{LHC1}, we show the fragmentation cross sections for UPCs of 
${}^{238}\text{U}$ (filled blue circles) and ${}^{208}\text{Pb}$ (open      
red squares) nuclei at the LHC. The mass distribution of Pb fragments displays 
a strong decrease of the distribution as the mass of the fragment decreases,    
mainly due to neutron emission. The same behavior is observed for uranium ions, 
but a noticeable number of fission fragments originating from the excitation of  
the ${}^{238}\text{U}$ projectile produces the observed double-hump structure,  
which is a characteristic signature of fission, with pronounced peaks with masses 
around $A \sim 100$ and $A \sim 140$.

\begin{figure}[h]
\begin{center}
\includegraphics[scale=0.42]{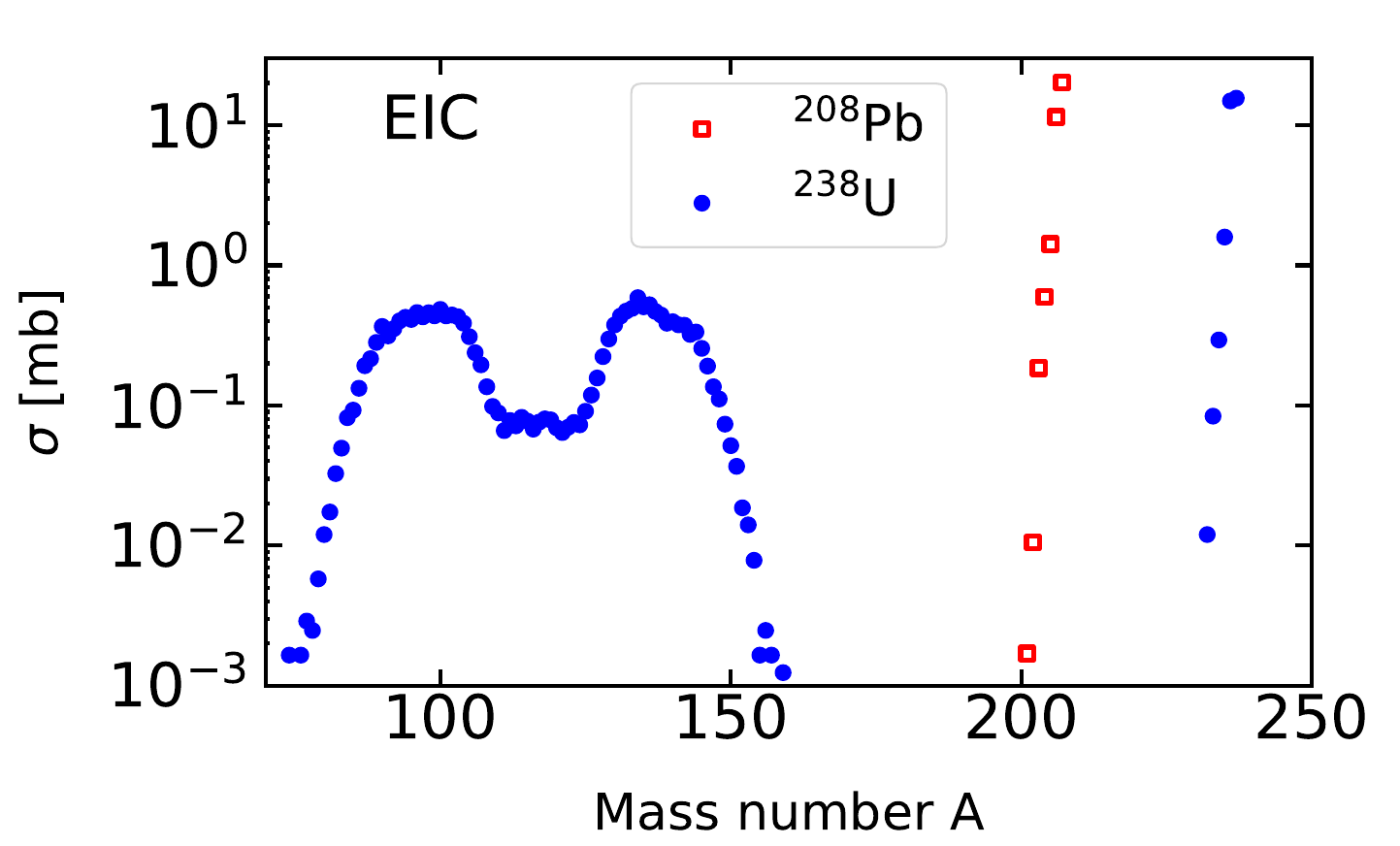}
\caption{Cross sections for the production of fragments as a function of their   
masses for electron-$^{208}$Pb (open red squares) and electron-$^{238}$U (filled
blue circles) collisions  at the LHC.  \label{EICws}}
\end{center}
\end{figure}

Fig. \ref{EICws} shows the fragmentation cross sections of ${}^{238}\text{U}$  
(filled blue circles) and ${}^{208}\text{Pb}$ (open red squares) nuclei at the 
EIC. The magnitude of the cross sections are about $10^3$ times smaller than at 
the LHC. The mass distributions of fragments for both nuclei also show a strong 
decrease of the cross sections with decreasing mass number of the fragment due to 
neutron emission. Again, the peaks of fission fragments around $A \sim 100$ and 
$A \sim 140$ for uranium nuclei are visible. It is evident that neutron removal 
from both nuclei decreases much faster than for UPCs at the LHC. Almost all events 
go to evaporation of a few neutrons and fission fragments.

The behavior displayed in Figs. \ref{LHC1} and \ref{EICws} is better understood  
with numerical values. We show this in Table \ref{tab1} with cross sections for 
neutron evaporation ($\sigma_{-1n}$, $\sigma_{-2n}$, $\sigma_{-3n}$, and    
$\sigma_{-4n}$), fission, total cross sections, and branching ratios to fission  
channels. The cross sections are given in barns at the LHC and in milibarns at the 
EIC, and it is clear that at the LHC the number of events in these channels is    
about a factor of $10^3$ larger than at the EIC. The isotope distribution in both 
cases show a dominance of neutron emission. Some fragments originate from the 
emission of light nuclei and the mass distribution drops dramatically. For 
$^{238}$U, fission fragments are produced in both laboratories.

At the EIC, the integrated luminosities for electron-nucleus collisions are expected to range between $10^{33}-10^{34}$ cm$^{2}$s$^{-1}$. 
Assuming a luminosity of ${\cal L} = 10^{34}$ cm$^{2}$s$^{-1}$, approximately  $10^{8}$ neutrons per second can be produced from electron-uranium nucleus fragmentation. Event rates for the production of uranium fission fragments are anticipated to fall within a similar range. These neutrons and fragments will predominantly travel in the forward direction with negligible transverse momenta, making their detection challenging without specialized equipment. A possible solution is to utilize a setup similar to the Zero Degree Calorimeter (ZDC) employed with the ALICE detector at CERN  \cite{ARNALDI2006235}. The specifics of detection techniques, however, lie beyond the scope of this work \cite{goto19}. Notably, many low-energy processes in relativistic heavy-ion collisions were predicted as early as the 1980s \cite{BERTULANI1988299} when suitable detectors were unavailable, yet they were eventually observed experimentally decades later. It is highly plausible that the low-energy processes described in this study will also become experimentally accessible at the EIC.

\begin{figure}[h]
\begin{center}
\includegraphics[scale=0.42]{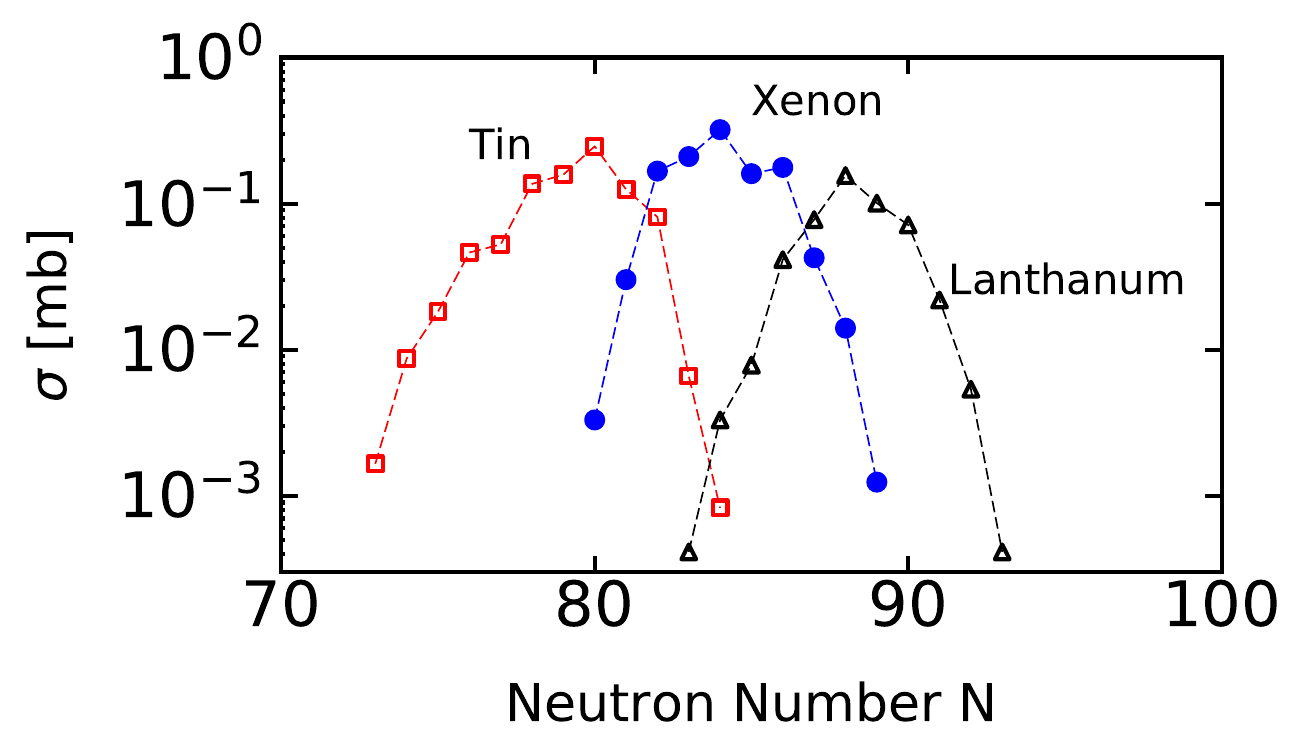}
\caption{Cross sections for the production of tin (open red squares), xenon    
(filled blue circles) and neodymium (open black triangles) isotopes at the EIC.
\label{EIC2}}
\end{center}
\end{figure}

The isotope production cross sections for tin (open red squares), xenon          
(filled blue circles), and lanthanum (open black triangles) at the EIC are shown 
in Fig. \ref{EIC2}. The electromagnetic interaction induces a small excitation    
energy of the nucleus - around 15 MeV -  compared to the large excitation due to strong interactions \cite{PhysRevLett.124.132301}. Because of this small excitation energy, the cross 
sections for the production of neutron-rich isotopes are much smaller than those 
for stable isotopes, as displayed in Fig. \ref{EIC2}. 
Strong lepton-nucleus interactions can also be treated in terms of initial       
lepton-parton hard scattering (primary interaction) followed by an intra-nuclear  
cascade collision model, as described in \cite{magdy2024}. This leads to the 
production of numerous neutron-deficient isotopes in the $Z=89-94$ region  \cite{Schmookler2023}. However, as remarked earlier, the cross sections are  
smaller than those calculated here because the initial lepton-parton cross   
section is smaller than the collective excitation of giant resonances, which 
yields a very large nuclear response to the electromagnetic field.

Our results suggest that the study of nuclear fragmentation in UPCs at the LHC  
might be better suited for nuclear physics investigations, provided that proper  
experimental conditions are met. The EIC has the advantage that the electron 
is a cleaner probe of the nucleus and that the fragments might be more easily 
identified.
\begin{center}
  \begin{table}[h]
 \begin{tabular}{|c|c|c|c|c|}
 \hline
 Cross sections     & LHC & LHC & EIC &EIC  \\
 \hline
  & Pb + Pb [b]& U + U [b] & e-Pb [mb] &e-U [mb]  \\
    \hline
    \hline
    $\sigma_{-1n}$ & 33.93 & 33.20&  20.24& 15.58\\
    $\sigma_{-2n}$ & 18.89 &30.59&  11.45& 14.88\\
    $\sigma_{-3n}$ & 2.546 &3.537&  1.416& 1.591\\
     $\sigma_{-4n}$ & 1.091 &0.784&  0.5933& 0.2934\\
     $\sigma_{fission}$ & 0 &18.24&  0& 8.867\\
     \hline
     $\sigma_{total}$ & 55.74 &85.48&  33.90& 41.32 \\  
     \hline  
Fission b.r.& 0\% &19.54\%&  0\%& 21.45\%\\   
\hline
    \end{tabular}
\caption{Cross sections (in barns at the LHC and milibarns at the EIC) for numerous  
decay channels of lead and uranium nuclei in collisions at the LHC and at the future 
EIC. Neutron (-1n, -2n , -3n and -4n) evaporation and fission channels are shown, as 
well as the total cross sections and the percentage of fission branching channels. }
    \label{tab1}
  \end{table}
\end{center}
{\it Conclusions.} 
We have compared low-energy nuclear excitation in Ultraperipheral 
Collisions at the LHC with $eA$ scattering at the future EIC. Our study confirms the applicability 
of the equivalent photon method for the energy regime of the EIC.
In UPCs at the LHC, careful experimental separation of purely electromagnetic 
interactions from hadronic interactions is required. In contrast, at the EIC, 
nuclear fragmentation is predominantly driven by electromagnetic interactions, 
providing a clearer probe of nuclear excitation followed by particle emission or 
nuclear fission. 

We applied our theoretical framework to describe the nuclear fragmentation of 
$^{208}$Pb and $^{238}$U ions at the EIC and in UPCs at the LHC. Our results 
indicate that the cross sections and number of events at the EIC are smaller than 
at the LHC by about a factor $10^3$. We considered only electromagnetic interactions
 through the excitation of giant resonances, which predominantly decay via neutron 
emission and, for $^{238}$U, by fission. Most isotopes produced are near the 
stability line due to the small excitation energy imparted by the electromagnetic 
interaction. Notably, about 21\% of events involving fragmentation of $^{238}$U at 
the EIC result in a large number of isotopes with masses around $A = 100$ and 
$A = 140$. The cross sections for neutron emission are so significant that they can 
be utilized for monitoring the degradation of the ion beam at the EIC.

At the EIC, the contribution of initial photo-nucleon interactions followed by parton-parton or nucleon-nucleon
cascades results in the excitation and decay of the nucleus. But their cross sections are small compared 
to the collective excitation of giant resonances. In heavy ion colliders at the 
LHC, nuclear fragmentation can also arise from central collisions, potentially 
forming a fireball with a QCD phase transition and subsequent hadronization. Our 
comparison of electromagnetic fragmentation at the EIC and the LHC highlights two 
extreme conditions for such processes and also underscores the potential of 
complementary studies at future high-power laser facilities to explore similar 
phenomena \cite{negoita2022,PhysRevAccelBeams.25.101601}.

In contrast to $eA$ collisions at the EIC, in UPCs we may have pomeron 
exchange, which is a long range strong interaction and competes with photon 
induced processes. In principle, pomerons can also cause nuclear fragmentation. There are also clear differences between the two processes as
pomerons are $0^{++}$ particles and the photon is $1^{--}$. In the EIC all processes are 
free of pomerons and it is an ideal laboratory to isolate pure electromagnetic from strong interactions processes in studies of low excitation energy nuclear physics.

\bigskip

{\it Acknowledgement.}  This work has been supported by the Turkish Council of Higher Education (YOK) under Mevlana Exchange Program, and by the U.S. DOE grants DE-FG02-08ER41533 and  the U.S. NSF Grant No. 1415656.



\end{document}